%% file: trujillo-bueno.tex
\documentclass{svmult}
\usepackage{epsfig,graphicx} 
\usepackage[bottom]{footmisc}
\usepackage{natbib,url}      
\bibpunct{(}{)}{;}{a}{}{,}   
\def\cite#1{\citealp{#1}}    
\def\authorindex#1{}  
\def\figspath{.}  

\begin{document}\newcount\preprintheader\preprintheader=1

\input{rr-assp-defs}

\title*{Recent Advances in Chromospheric and Coronal Polarization 
        Diagnostics}

\titlerunning{Chromospheric and Coronal Polarization Diagnostics}

\author{J. Trujillo Bueno \inst{1,2}}

\authorindex{Trujillo Bueno, J.}
 
\institute{
   Instituto de Astrof\'\i sica de Canarias, La Laguna, Tenerife,
   Spain \and Consejo Superior de Investigaciones Cient\'\i ficas,
   Spain}

\maketitle

\setcounter{footnote}{0}  

\begin{abstract} 
  I review some recent advances in methods to diagnose polarized
  radiation with which we may hope to explore the magnetism of the
  solar chromosphere and corona.  These methods are based on the
  remarkable signatures that the radiatively induced quantum
  coherences produce in the emergent spectral line polarization and on
  the joint action of the Hanle and Zeeman effects. Some applications
  to spicules, prominences, active region filaments, emerging flux
  regions and the quiet chromosphere are discussed.
\end{abstract}

\section{Introduction}      \label{jtb-introduction}

The fact that the anisotropic illumination of the atoms in the
chromosphere and corona induces population imbalances and quantum
coherences between the magnetic sublevels, even among those pertaning
to different levels, is often considered as a hurdle for the
development of practical diagnostic tools of ``measuring'' the
magnetic field in such outer regions of the solar atmosphere. However,
as we shall see throughout this paper, it is precisely this fact that
gives us the hope of reaching such an important scientific goal. The
price to be paid is that we need to develop high-sensitivity
spectropolarimeters for ground-based and space telescopes and to
interpret the observations within the framework of the quantum theory
of spectral line formation. As J.W.~Harvey put it, ``this is a hard
research area that is not for the timid'' (\cite{harvey-review}).

Rather than attempting to survey all of the literature on the subject,
I have opted for beginning with a very brief introduction to the
physics of spectral line polarization, pointing out the advantages and
disadvantages of the Hanle and Zeeman effects as diagnostic tools, and
continuing with a more detailed discussion of selected
developments. Recent reviews where the reader finds complementary
information are \citet{harvey-review}, \citet{stenflo-spw4},
\citet{lagg-review},
\citet{lopez-ariste-review}, \citet{casini-landi-review}, and
\citet{trujillo-spw5}.

\section{The physical origin of the spectral line polarization}
\label{jtb-sec2}

Solar magnetic fields leave their fingerprints in the polarization
signatures of the emergent spectral line radiation. This occurs
through a variety of rather unfamiliar physical mechanisms, not only
via the Zeeman effect. In particular, in stellar atmospheres there is
a more fundamental mechanism producing polarization in spectral
lines. There, where the emitted radiation can escape through the
stellar surface, the atoms are illuminated by an anisotropic
radiation field. The ensuing radiation pumping produces population
imbalances among the magnetic substates of the energy levels (that is,
atomic level polarization), in such a way that the populations of
substates with different values of $|M|$ are different ($M$ being the magnetic quantum number). 
This is termed {\em atomic level alignment\/}.
As a result, the emission process can
generate linear polarization in spectral lines without the need for a
magnetic field. This is known as scattering line polarization (e.g.,
\cite{stenflo-book}, \cite{landi-landolfi}). Moreover, radiation is also selectively absorbed
when the lower level of the transition is polarized
(\cite{trujillo-landi-97}, \cite{trujillo-spw2},
\cite{trujillo-nature02}, \cite{manso-trujillo-prl}). Thus, the medium
becomes dichroic simply because the light itself has the chance of
escaping from it.

Upper-level polarization produces {\em selective emission\/} of
polarization components, while lower-level polarization produces {\em
selective absorption\/} of polarization components. A useful
expression to estimate the amplitude of the emergent fractional linear
polarization is the following generalization of the Eddington-Barbier
formula (\cite{trujillo-spw3}), which establishes that the emergent
$Q/I$ at the center of a sufficiently strong spectral line when
observing along a line of sight (LOS) specified by 
$\mu=\cos \theta$ (with $\theta$ being the angle between the local solar vertical and the LOS) 
is approximately given by
\begin{equation}
{{Q}\over{I}}\,{\approx}\,{3\over{2\sqrt{2}}}(1-\mu^2)\,[{\cal W}\,\sigma^2_0({J_u})\,-\,{\cal Z}\,\sigma^2_0({J_l})]\,=\,{3\over{2\sqrt{2}}}(1-\mu^2)\,\,{\cal F},
\end{equation}
where ${\cal W}$ and ${\cal Z}$ are numerical factors that depend on
the angular momentum values ($J$) of the lower ($l$) and upper ($u$)
levels of the transition (e.g., ${\cal W}={\cal Z}=-1/2$ for a line
with $J_l=J_u=1$), while $\sigma^2_0=\rho^2_0/\rho^0_0$ quantifies the
fractional atomic alignment of the upper or lower level of the
spectral line under consideration, calculated in a reference system
whose $Z$-axis (i.e., the quantization axis of total angular momentum)
is along the local solar vertical.\footnote{For example,
$\rho^0_0(J=1)=(N_1+N_0+N_{-1})/\sqrt{3}$ and
$\rho^2_0(J=1)=(N_1-2N_0+N_{-1})/\sqrt{6}$, where $N_1$, $N_0$ and
$N_{-1}$ are the populations of the magnetic sublevels.} The
$\sigma^2_0(J)$ values quantify the degree of population imbalances
among the sublevels of level $J$ with different $|M|$-values.  They
have to be calculated by solving the statistical equilibrium equations
for the multipolar components of the atomic density matrix (see
Chapt.~7 of \cite{landi-landolfi}).
In a weakly anisotropic medium like the solar atmosphere, the
$\sigma^2_0(J_l)$ and $\sigma^2_0(J_u)$ values of a resonance line
transition are proportional to the so-called anisotropy factor
$w=\sqrt{2}J^2_0/J^0_0$ (e.g., \S3 in \cite{trujillo-01}), where
$J^0_0$ is the familiar mean intensity and ${J}^2_0 \approx \oint {\rm
d} \vec{\Omega}/(4\pi)\,1/(2\sqrt{2})\,(3\mu^2-1)\,{{I_{\nu,
\vec{\Omega}}}}\,$ quantifies whether the illumination of the atomic
system is preferentially vertical ($w>0$) or horizontal ($w<0$). Note
that in Eq.~(1) the $\sigma^2_0$ values are those corresponding to the
atmospheric height where the line-center optical depth is unity along
the LOS.

The most practical aspect is that a magnetic field inclined with
respect to the symmetry axis of the pumping radiation field modifies
the atomic level polarization via the Hanle effect (e.g., the reviews
by \cite{trujillo-01}, \cite{trujillo-leuven}; see also
\cite{landi-landolfi}). Approximately, the amplitude of the emergent
spectral line polarization is sensitive to magnetic strengths between
$0.1\,B_{\rm H}$
and $10\,B_{\rm H}$, where the critical Hanle field intensity ($B_{\rm
H}$, in gauss) is that for which the Zeeman splitting of the $J$-level
under consideration is equal to its natural width:
\begin{equation} 
B_{\rm H} = 1.137{\times}10^{-7}/(t_{\rm life}\,g_J)
\end{equation}
with $t_{\rm life}$ the lifetime, in seconds, of the $J$-level under
consideration and $g_J$ its Land\'e factor.  Since the lifetimes
of the upper levels ($J_u$) of the transitions of interest are usually
much smaller than those of the lower levels ($J_l$), clearly diagnostic
techniques based on the lower-level Hanle effect are sensitive to much
weaker fields than those based on the upper-level Hanle effect.

The Hanle effect gives rise to a rather complex magnetic-field
dependence of the linear polarization of the emergent spectral line
radiation. In the saturation regime of the upper-level Hanle effect
(i.e., when the magnetic strength $B>B_{\rm satur} \approx 10\,B_{\rm
H}(J_u)$, with $B_{\rm H}(J_u)$ the critical Hanle field of the line's
upper level) it is possible to obtain manageable formulae for the
line-center amplitudes of the emergent linear polarization profiles,
which show that in such a regime the $Q/I$ and $U/I$ signals only
depend on the orientation of the magnetic field vector. Assume, for
simplicity, a deterministic magnetic field with $B>B_{\rm satur}$
inclined by an angle $\theta_B$ with respect to the local solar
vertical (i.e., the $Z$-axis) and contained in the $Z$--$Y$ plane.
Consider any LOS contained in the $Z$--$X$ plane, characterized by
$\mu=\cos \theta$. Choose the $Y$-axis direction as the reference
direction for Stokes $Q$.  It can be shown that the following
approximate expressions hold for the emergent linear polarization
amplitudes in an electric-dipole transition\footnote{For magnetic
dipole transitions it is only necessary to change the sign of the
$Q/I$ and $U/I$ expressions given in this paper. To understand the
reason for this see \S6.8 of \citet{landi-landolfi}.}:
\begin{equation} 
{{Q}\over{I}}\,{\approx}\,{{3}\over{8\sqrt{2}}}\,{\Big[}(1-\mu^2)(3\cos^2{\theta_B}\,-\,1)\,+\,(1+\mu^2)(\cos^2{\theta_B}\,-\,1){\Big]}\,(3\cos^2{\theta_B}\,-\,1)\,\,{\cal
F},
\end{equation}
\begin{equation} 
{{U}\over{I}}\,{\approx}\,-{{3}\over{2\sqrt{2}}}\,\sqrt{1-\mu^2}\,\sin {\theta_B}\cos{\theta_B}\,(3\cos^2{\theta_B}\,-\,1)\,\,{\cal F},
\end{equation}
where ${\cal F}={\cal W}\,\sigma^2_0\,({J_u})\,-\,{\cal
Z}\,\sigma^2_0({J_l})$ is identical to that of Eq.~(1), which depends
on the $\sigma^2_0$ values for the {\em unmagnetized\/} reference case.

It is of interest to consider the following particular cases, ignoring
for the moment that in a stellar atmosphere the ${\cal F}$ value tends
to be the larger the smaller $\mu$. First, the $B=0$ case of Eq.~(1)
can be easily recovered by chosing $\theta_B=0^{\circ}$ in Eqs. (3)
and (4), because there is no Hanle effect if the magnetic field is
parallel to the symmetry axis of the incident radiation field. Second,
note that for $\theta_B=90^{\circ}$ (horizontal magnetic field)
$U/I=0$ and that for this case we find exactly the same $Q/I$
amplitude for all LOSs contained in the $Z$--$X$ plane, including that
with $\mu=1$ which corresponds to forward-scattering geometry. Note
also that Eq.~(3) implies that the amplitude of the forward-scattering
$Q/I$ signal created by the Hanle effect of a horizontal magnetic
field with a strength in the saturation regime is only a factor two
smaller than the $Q/I$ signal of the unmagnetized reference case in
$90^{\circ}$ scattering geometry (i.e., the case of a LOS with
$\mu=0$). Some interesting examples of detailed numerical calculations
of the emergent $Q/I$ and $U/I$ amplitudes for a variety of $\theta_B$
and $\mu$ values can be seen in Fig.~9 of \citet{hazel}, which the
reader will find useful to inspect.  Such results for the lines of the
\HeI~10830~\AA\ multiplet can be easily understood via Eqs. (3) and
(4).

It is easy to generalize Eqs.\ (3) and (4) for any magnetic field
azimuth $\chi_B$. Such general equations show clearly that there are
two particular scattering geometries (i.e., those with $\mu=0$ and
$\mu=1$) for which the Stokes profiles corresponding to
$\theta_B^{*}=180^\circ-\theta_B$ and $\chi_B^{*}=-\chi_B$ are
identical to those for which the magnetic field vector has $\theta_B$
and $\chi_B$ (i.e., the familiar ambiguity of the Hanle effect). If
the observed plasma structure is not located in the plane of the sky,
or if it is outside the solar disk center, one then has a
quasi-degeneracy which can disappear when $\mu$ is considerably
different from 1 or from 0. This fact can be exploited for removing
the $180^{\circ}$ azimuth ambiguity present in vector magnetograms
(\cite{landi-bommier-93}; see also Fig.~\ref{jtb-fig:atompol} below).

For the case of a magnetic field with a fixed inclination $\theta_B$
and a random azimuth below the spatial scale of the mean free path of
the line photons we have $U/I=0$, while
\begin{equation} 
  {{Q}\over{I}}\,{\approx}\,{{3}\over{8\sqrt{2}}}\,(1-\mu^2)\,
  [3\cos^2{\theta_B}\,-\,1]^{2}\,{\cal F}.
\end{equation}
This expression shows that under such circumstances there is no
forward scattering polarization. It shows also that the $Q/I$
amplitude of a scattering signal produced in the presence of a
horizontal magnetic field with a random azimuth and $B>B_{\rm satur}$
is a factor 4 smaller than $[Q/I]_{0}$ (i.e., than the $Q/I$ amplitude
corresponding to the unmagnetized reference case). It is also possible
to show that $Q/I\,{\approx}\,(1/5)\,[Q/I]_{0}$ for the case of a
microturbulent magnetic field with an isotropic distribution of field
directions.

Finally, it is important to emphasize that a rigorous modeling of the
polarization produced by the joint action of the Hanle and Zeeman
effects in many spectral lines of diagnostic interest requires
calculating the wavelength positions and the strengths of the $\pi$
and $\sigma$ components within the framework of the Paschen-Back
effect theory. This theory allows us to model the important
level-crossing regime in which the energy eigenvectors are gradually
evolving from the form ${\vert} L S J M \rangle$ (with $M$ the
projection of the total angular momentum $J=L+S$ along the
quantization axis) to the form ${\vert} L S M_L M_S \rangle$ as the
magnetic field increases. This range between the limiting cases of
``weak'' fields (Zeeman effect regime) and ``strong'' fields (complete
Paschen-Back regime) is called the incomplete Paschen-Back effect
regime. The reason why it is so important for a correct modeling of
the spectral line polarization in fine-structured and in
hyperfine-structured multiplets is because the level crossings and
repulsions that take place in this regime give rise to subtle
modifications of the atomic level polarization and, therefore, to a
number of remarkable effects on the emergent spectral line
polarization (e.g., \cite{bommier-80}, \cite{landi-82},
\cite{trujillo-sodium}, \cite{belluzzi-07}). Of particular interest is
the so-called alignment-to-orientation transfer mechanism studied by
\citet{landi-82} for the \HeI~ D$_3$ multiplet, by means of which a
fraction of the atomic level alignment produced by anisotropic pumping
processes can lead to {\em atomic level orientation\/} (i.e., an
atomic excitation situation such that the populations of substates
with magnetic quantum numbers $M$ and $-M$ are different). Obviously,
the observational signature of the presence of a significant amount of
atomic level orientation is a Stokes $V(\lambda)$ profile dominated by
one of its lobes.

\section{The Zeeman effect vs. the Hanle effect}

Good news is that the mere detection of Zeeman polarization
signature(s) implies the presence of a magnetic field. One
disadvantage of the polarization of the Zeeman effect as a diagnostic
tool is that it is blind to magnetic fields that are tangled on scales
too small to be resolved. Another drawback is that it is of limited
practical interest for the determination of magnetic fields in hot
(chromospheric and coronal) plasmas because the circular polarization
induced by the longitudinal Zeeman effect scales with the ratio,
${\cal R}$, between the Zeeman splitting and the Doppler width (which
is much larger than the natural width of the atomic
levels!). Likewise, given that for not too strong fields the Stokes
$Q$ and $U$ signals produced by the transverse Zeeman effect scale as
${\cal R}^2$, their amplitudes are normally below the noise level of
present observational possibilities for intrinsically weak fields
(typically, $B \lesssim 100$~gauss in solar spectropolarimetry).

The Hanle effect is especially sensitive to magnetic fields for which
the Zeeman splitting is comparable to the natural width of the upper
(or lower) level of the spectral line used, regardless of how large
the line width due to Doppler broadening is. Therefore, it is
sensitive to weaker magnetic fields than the Zeeman effect: from at
least 1~mG to a few hundred gauss (see Eq.~2). Moreover, it is
sensitive to magnetic fields that are tangled on scales too small to
be resolved (e.g., \cite{stenflo-book},
\cite{trujillo-nature04}). Finally, note that the diagnostic use of
the Hanle effect is {\em not\/} limited to a narrow solar limb
zone. In particular, in forward scattering at the solar disk center,
the Hanle effect can create linear polarization in the presence of
inclined magnetic fields (\cite{trujillo-nature02}). The disadvantage
is that the Hanle effect signal saturates for magnetic strengths $B >
10\,B_{\rm H}(J_u)$, a regime where the linear polarization signals are
sensitive only to the orientation of the magnetic field vector.

Fortunately, both effects can be suitably complemented for exploring
magnetic fields in solar and stellar physics.

\section{Diagnostic tools based on spectral line polarization}
\label{jtb-sec3}

The determination of the magnetic, dynamic and thermal properties of
solar plasma structures via the interpretation of the observed Stokes
profiles requires the development of suitable diagnostic tools. The
aim is to find the physical properties of the adopted model such that
the difference between the synthetic and the observed Stokes profiles
is the smallest possible one. Depending on the observed plasma
structure, the model chosen to represent it can be rather simple
(e.g., a constant-property slab) or more sophisticated (e.g., a
stratified, one-dimensional atmosphere model). The first step is to
develop an efficient way to compute the emergent Stokes profiles for
any plausible realization of the model's physical properties. Such
spectral synthesis tools can be used for doing forward modeling
calculations (e.g., in snapshot models taken from MHD simulations) or
for developing inversion codes of Stokes profiles induced by various
physical mechanisms. At present there are two Stokes inversion
approaches. One employs searching in databases of theoretical Stokes
profiles computed with the spectral synthesis tool, ideally for all
possible configurations of the model's physical properties. The other
employs iterative algorithms aiming to minimize the merit function
used to quantify the goodness of the fit of the model properties.
This is done by combining the spectral synthesis tool with suitable
minimization algorithms, such as the Levenberg-Marquardt method.

\subsection{Methods for the chromosphere and transition region}\label{sub2-1}

The intensity and polarization of the spectral lines that originate in
the bulk of the solar chromosphere (e.g., the IR triplet and the
K-line of \CaII) 
and in the transition region (e.g., Ly\,$\alpha$
and \MgII\,k) contain precious information on these
atmospheric regions.  In general, their linear polarization is due to
the joint action of the atomic level polarization and the Hanle and
transverse Zeeman effects, while their circular polarization is
dominated by the longitudinal Zeeman effect.

In regions with high concentrations of magnetic flux, such as in
sunspots, the polarization signals are dominated by the Zeeman
effect. Therefore, diagnostic techniques based on this effect are
quite useful. For example, the non-LTE inversion code of Stokes
profiles induced by the Zeeman effect developed by \citet{sn-tb-rc-00}
has led to several interesting applications (e.g.,
\cite{sn-tb-rc-science}, \cite{sn-sunspot-05},
\cite{pietarila-07-inv}).

In order to model spectropolarimetric observations of chromospheric
and transition region lines outside sunspots, 
it is necessary to take into account the atomic polarization
that anisotropic radiation pumping processes induce in the atomic
levels. This requires solving a significantly more complicated
radiative transfer problem, known as the non-LTE problem of the
$2^{\rm nd}$ kind (see \cite{landi-landolfi}). It consists in
calculating, at each spatial point of any given atmospheric model and
for each $J$-level of the chosen atomic model, the density matrix
elements that are consistent with the intensity and
polarization of the radiation field generated within the (generally
magnetized) medium under consideration. Once such density matrix
elements are known it is possible to solve the Stokes vector transfer
equation for any desired LOS with an accurate and efficient formal
solution method, such as the DELOPAR technique discussed by
\citet{trujillo-tubinga}.

To that end, \citet{manso-trujillo-spw3} developed MULTIPOL, a general
radiative transfer computer program for solving multilevel scattering
polarization problems including the Hanle and Zeeman effects of a weak
magnetic field (see also \cite{manso-thesis}). MULTIPOL is based on
the multilevel atom model of the quantum theory of spectral line
formation (see \S7.2 in \cite{landi-landolfi}), which allows us to
take into account that the mean intensity and anisotropy of the
various line transitions pertaining to any given multiplet can be
different. A similar spectral synthesis code based also on the DELOPAR
technique and on the iterative scheme proposed by
\citet{trujillo-spw2} has been recently developed by
\citet{stepan-thesis}. Solving the ensuing Stokes inversion problem
for the magnetic field vector is possible, but requires to adopt
a model (or a few plausible models) for the thermal and density stratifications.
 
The quantum theory of spectral line polarization on which the
above-mentioned computer programs are based treats the scattering line
polarization phenomenon as the temporal succession of 1st-order
absorption and re-emission processes, interpreted as statistically
independent events (complete redistribution in frequency). This theory
is very suitable for modeling the polarization observed in many
diagnostically important spectral lines, such as the IR triplet of
\CaII\ and \Halpha. It can also be used for estimating the
line-center polarization amplitude in lines for which frequency
correlations between the incoming and outgoing photons are significant
(e.g., \CaIIK, Ly\,$\alpha$, \MgII\,k), especially in
forward-scattering geometry at the solar disk center. However, for
interpreting particular spectral features that are observed in the
wings of some strong lines it is necessary to apply a theory not based
on the Markov approximation. At present, such a formulation is only
available for the particular case of a two-level atom without
lower-level polarization (e.g., \cite{sampoorna-07} and references
therein). Fortunately, the lower level of several resonance line
transitions cannot be aligned, so that modeling efforts in this
direction are of interest (e.g., \cite{holzreuter-stenflo-K}).

\subsection{Methods for chromospheric and coronal structures}\label{sub2-2}

There are several possibilities for determining the magnetic field
vector that confines and/or channels the plasma of structures embedded
in the optically thin outer layers of the solar atmosphere, such as
prominences, spicules, active region filaments, etc. At present, the
best available option is to choose spectral lines entirely produced by
the plasma structures themselves, such as those of the \HeI~10830~\AA\
and 5876 \AA\ (D$_3$) multiplets, and to interpret observations of
their intensity and polarization within the framework of the multiterm
atom model of the quantum theory of spectral line formation (see \S7.6
in \cite{landi-landolfi}). The spectral lines of the
\HeI~10830~\AA\ and 5876 \AA\ multiplets result from transitions
between terms of the triplet system of helium (ortho-helium), whose
respective $J$-levels (with $J$ the level's total angular momentum)
are far less populated than the ground level of helium (the singlet
level $^1$S$_0$). The lower term (2s$^3$S$_1$) of the
\HeI~10830~\AA\ multiplet is the ground level of ortho-helium, while
its upper term (2p$^3$P$_{2,1,0}$) is the lower one of 5876 \AA\
(whose upper term is 3d$^3$D$_{3,2,1}$).

The Stokes profiles of the \HeI~10830~\AA\ and 5876 \AA\ multiplets
depend on the strengths and wavelength positions of the $\pi$ and
$\sigma$ transitions, which can only be calculated correctly within
the framework of the Paschen-Back theory. In fact, the
sublevels of the $J=2$ and $J=1$ upper levels of the \HeI~10830~\AA\
triplet cross between 400~G and 1600~G, approximately, while the
sublevels of the $J=3$ and $J=2$ upper levels of the \HeI~5876~\AA\
multiplet show several crossings for field strengths of the order of 10~G
(e.g., Fig.~3 of \cite{hazel}). Moreover, the emergent Stokes
profiles can be seriously affected by the presence of atomic level
polarization produced by anisotropic radiative pumping processes,
which can be very significant even in the metastable (long-lived)
lower level of the \HeI~10830~\AA\ multiplet
(\cite{trujillo-nature02}). Elastic collisions with the neutral
hydrogen atoms of the solar chromospheric and coronal structures are
unable to destroy the atomic polarization of the \HeI~levels.

It is important to put reliable codes for the synthesis and inversion
of Stokes profiles at the disposal of the astrophysical community. To
this end, \citet{hazel} developed a user-friendly computer program
called HAZEL (from HAnle and ZEeman Light), which takes into account
all the relevant physical mechanisms and ingredients (optical pumping,
atomic level polarization, level crossings and repulsions, Zeeman,
Paschen-Back and Hanle effects). The user can either calculate the
emergent intensity and polarization for any given magnetic field
vector, or can infer the dynamical and magnetic properties from the
observed Stokes profiles via an efficient inversion algorithm based on
global optimization methods. The influence of radiative transfer on
the emergent spectral line radiation is taken into account through a
suitable constant-property slab model, in which the
radiatively-induced atomic level polarization is assumed to be
dominated by the photospheric continuum radiation. At each point of
the observed field of view the slab's optical thickness is chosen to
fit the observed Stokes $I$ profile, a strategy which accounts
implicitly for the true physical mechanisms that populate the \HeI\ triplet
levels (e.g., the photoionization-recombination mechanism
discussed by \cite{avrett}, \cite{centeno-uv}, and others). The
observed Stokes $Q$, $U$ and $V$ profiles are then used to infer the
magnetic field vector.

\begin{figure}  
  \centering
  \includegraphics[width=10cm]{\figspath/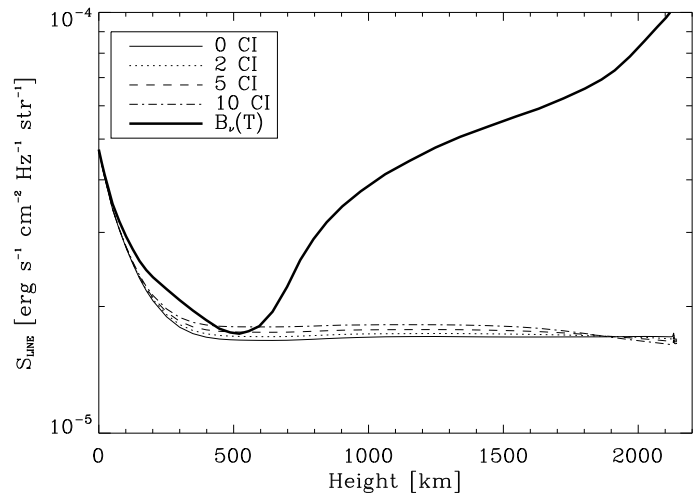}
\caption[]{\label{jtb-fig:he-uv}
  The variation with height in a semi-empirical solar-atmosphere model
  of the line source function of the \HeI~10830~\AA\ transitions,
  calculated from the non-LTE population values given in Fig.~6 of
  \citet{centeno-uv}, which indicates that significant absorption of
  the photospheric radiation at 10830 \AA\ is to be expected
  mainly around 2000 km in the atmosphere model.  Each curve
  corresponds to the indicated value of the EUV coronal irradiance
  (CI), in units of the nominal value.  }
\end{figure}

It is important to clarify that the assumption of a constant line
source function within the slab is reasonable for the
\HeI~10830~\AA\ and D$_3$ multiplets, as can be deduced from 
non-LTE calculations of the populations of the \HeI~levels
that take into account the influence of the EUV radiation that
penetrates the chromosphere from the overlying corona (e.g., Fig.~6 of
\cite{centeno-uv}). This can be seen more clearly in
Fig.~\ref{jtb-fig:he-uv}, which shows the height variation of the line
source function in any of the \HeI~10830~\AA\ transitions for
increasing values of the EUV irradiance. \citet{avrett} find also that
the line source function of the \HeI~10830~\AA\ triplet is essentially
constant, and equal to about $0.4{\times}I_c$ (with $I_c$ the
continuum intensity at 10830 \AA).

\begin{figure}  
  \centering
  \includegraphics[width=\textwidth]{\figspath/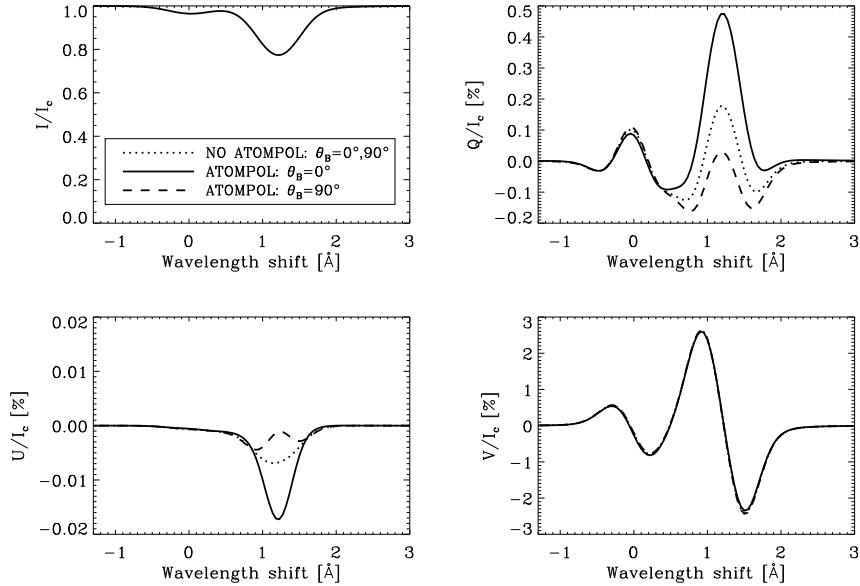}
  \caption[]{\label{jtb-fig:atompol} The emergent Stokes profiles in
  the \HeI~10830~\AA\ triplet calculated with the spectral synthesis
  option of HAZEL for a LOS with $\theta=45^{\circ}$ ($\mu=0.707$)
  contained in the $Z$--$X$ plane, with the $Z$-axis along the local solar
  vertical. The positive reference direction for Stokes $Q$ is along
  the $Y$-axis. The calculations assume a magnetized ($B=1200$~G!) slab
  of helium atoms located at a height of about 2200~km 
  above the visible solar surface. Two magnetic field orientations are
  considered: ($a$) vertical field along the $Z$-axis
  ($\theta_B=0^{\circ}$ and $\chi_B=0^{\circ}$) and ($b$) horizontal
  field along the $X$-axis ($\theta_B=90^{\circ}$ and
  $\chi_B=0^{\circ}$). The slab's optical thickness at the wavelength
  of the red blended component is $\Delta{\tau}_{\rm red}=0.5$. The
  dotted curves show the emergent Stokes profiles for cases $a$ and $b$
  when only the Zeeman effect is taken into account. The other curves
  show the emergent Stokes profiles when taking into account also the
  influence of atomic level polarization in case $a$ (solid)
  and in case $b$ (dashed).
  Note that the influence of atomic level polarization on the
  linear polarization profiles is very significant and that it removes
  the $180^{\circ}$ azimuth ambiguity present in the Zeeman-effect
  profiles. Interestingly, the non-zero Stokes $U$ signals are caused
  by the anomalous dispersion terms.}
\end{figure}

Figure~\ref{jtb-fig:atompol} shows an example of a model calculation
carried out with the synthesis option of HAZEL. For the case of a
plasma structure levitating at a height of 2200~km above the
visible solar surface and permeated by a magnetic field of 1200~G, the
figure shows two types of calculations of the emergent Stokes profiles
in the \HeI~10830~\AA\ multiplet for a LOS with
$\theta=45^{\circ}$. The calculations have been carried out for the
two magnetic field orientations indicated in the figure legend: ($a$)
vertical field and ($b$) horizontal field. When the atomic level
polarization is neglected, the Zeeman effect caused by both
magnetic field vectors produce exactly the same Stokes profiles (see
the dotted lines). This is because the circular polarization of the
Zeeman effect depends on the inclination of the magnetic field vector
with respect to the LOS (which is identical for cases $a$ and $b$) and
the linear polarization is invariant when the component of the field
in the plane perpendicular to the LOS is rotated by $180^{\circ}$
(i.e., the well-known azimuth ambiguity of the Zeeman
effect). However, when the influence of atomic polarization is taken
into account then the linear polarization profiles corresponding to
such magnetic field orientations are very different (see the solid and
dashed curves), simply because in the presence of atomic level
polarization the emergent Stokes profiles not only depend on the
orientation of the magnetic field with respect to the LOS, but also on
its inclination with respect to the local solar vertical (e.g.,
\cite{landi-bommier-93}). In fact, the information provided by
Eqs. (3) and (4) is contained in the following single formula
(cf., \cite{trujillo-spw3})
\begin{equation}
  {{Q}\over{I}}\,\approx\,-\,{3\over{4\sqrt{2}}}\,\sin^2{\Theta}\,(3\cos^2{\theta_B}\,-\,1)\,{\cal
F},
\end{equation}
where $\Theta$ is the angle between the magnetic field vector and the
LOS, $\theta_B$ the angle between the magnetic field vector and the
local solar vertical, and the reference direction for Stokes $Q$ is
that for which $U/I=0$ (i.e., the parallel to the projection of the
magnetic field onto the plane perpendicular to the LOS). 
The sign of ${\cal F}$ is established by the solution of the
statistical equilibrium equations for the elements of the atomic
density matrix.  This formula shows clearly why the so-called
Van-Vleck angle, $\theta_{\rm V}={54.^{\circ}}74$, is {\em magic\/}.
Since $\cos^2 {\theta_{\rm V}}=1/3$, it is clear that $Q/I$ has the
sign of ${\cal F}$ for $\theta_{\rm V} < \theta_B < \pi -
\theta_{\rm V}$, but the opposite sign for $0 < \theta_B < \theta_{\rm
V}$ or $\pi - \theta_{\rm V} < \theta_B < \pi$.  Eq.~(6) can also be
used to understand what the Van-Vleck ambiguity of the Hanle effect is
(e.g., \cite{casini-judge}, \cite{ariste-casini-spicules},
\cite{merenda-06}).

Interestingly, as shown also in Fig.~\ref{jtb-fig:atompol}, for some
spectral lines, such as those of the \HeI~10830~\AA\ and D$_3$ multiplets,
the influence of atomic level alignment on the emergent linear
polarization can be very important, even in the presence of magnetic
fields as strong as 1200~G (\cite{trujillo-asensio-07}). Therefore,
inversion codes that neglect the influence of atomic level
polarization, such as the Milne-Eddington codes of \citet{lagg-04}
and \citet{sn-tb-ld-04}, should ideally be used only for the inversion
of Stokes profiles emerging from strongly magnetized regions (e.g.,
with $B>2000$~G for the case of the \HeI~10830~\AA\ triplet) or when
the observed Stokes $Q$ and $U$ profiles turn out to be dominated by
the transverse Zeeman effect, as happens with some active region
filaments (see \S6.2).\footnote{It is, however, important to note that
the positions and strengths of the $\pi$- and $\sigma$-components must
be calculated within the framework of the Paschen-Back effect theory, 
even in the Zeeman-dominated case (see \cite{sn-tb-ld-04}).}

The inversion option of HAZEL is based on the Levenberg-Marquardt (LM)
method for locating the minimum of the merit function that quantifies
the goodness of the fit between the observed and synthetic Stokes
profiles. In order to improve the convergence properties of the LM
method, HAZEL uses a novel initialization technique based on the
DIRECT algorithm, a deterministic global optimization technique
that is significantly more efficient than the stochastic method
PIKAIA considered by \citet{charbo} and used by \citet{lagg-04} in
their inversion code of Stokes profiles induced by the
Zeeman effect. This code, called H{\small E}LI{\small X}, has been recently 
improved by combining its inversion approach with the (Hanle+Zeeman) 
spectral synthesis calculation core of HAZEL (see \cite{helix}). 
An alternative inversion procedure is the Principal
Component Analysis (PCA) technique described by
\citet{ariste-casini-pca}, which necessitates first creating a suitable
database of emergent Stokes profiles for a comprehensive set of
illumination, thermodynamic, and magnetic conditions in the plasma
structure under consideration. Another inversion
strategy based on databases was applied by \citet{trujillo-spicules}
and \citet{merenda-06} to spectropolarimetric observations of solar
spicules and prominences in the \HeI~10830~\AA\ multiplet, which is of
particular interest for the determination of the magnetic field vector
in plasma structures with $10\,{\rm G} \lesssim \,B\, \lesssim 100\,{\rm G}$.

\section{The quiet chromosphere}\label{jtb-sec4}
 
A very suitable diagnostic window for mapping the magnetic fields of
the ``quiet'' regions of the solar chromosphere is that provided by the
polarization signals of the \CaII~IR triplet
(\cite{manso-trujillo-coimbra}). In such regions the circular
polarization of the \CaII~IR lines is caused by the longitudinal
Zeeman effect, while the Stokes $Q$ and $U$ profiles are dominated by
atomic level polarization and the Hanle effect. Interestingly, while
the linear polarization in the 8498 \AA\ line shows a strong
sensitivity to inclined magnetic fields with strengths between 1~mG
and 10~G, the emergent linear polarization in the 8542 \AA\ and 8662
\AA\ lines is very sensitive to magnetic fields in the milligauss
range. The reason for this very interesting behavior is that the
scattering polarization in the 8498 \AA\ line gets a significant
contribution from the selective emission processes that result from
the atomic polarization of the short-lived upper level, while that in
the 8542 \AA\ and 8662 \AA\ lines is dominated by the selective
absorption processes that result from the atomic polarization of the
metastable (long-lived) lower levels
(\cite{manso-trujillo-prl}). Therefore, in the quiet chromosphere the
magnetic sensitivity of the linear polarization of the 8542 \AA\ and
8662 \AA\ lines is mainly controlled by the lower-level Hanle effect,
which implies that in regions with $1\,{\rm G} \lesssim B \lesssim
50$~G the Stokes $Q$ and $U$ profiles are only sensitive to the
orientation of the magnetic field vector.  The 8498 \AA\ line is
however sensitive to both the orientation and the strength of the
magnetic field through the upper-level Hanle effect.

\begin{figure}  
  \centering
  \includegraphics[width=11.0cm]{\figspath/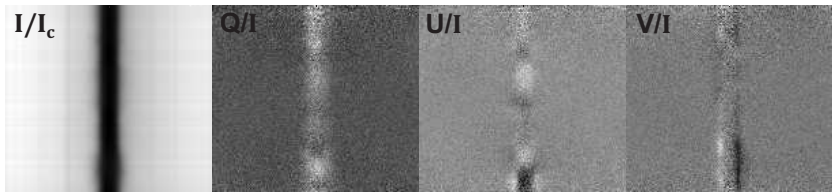}
\caption[]{
  An example of our recent spectropolarimetric observations of the
  \CaII~8542~\AA\ line in a very quiet region close to the solar
  limb, using ZIMPOL at the French-Italian telescope THEMIS. 
  The reference direction for Stokes
  $Q$ is the tangent to the closest limb. From
  \citet{trujillo-themis-zimpol}.
\label{jtb-fig:themis-zimpol}
}\end{figure}

Figure~\ref{jtb-fig:themis-zimpol} shows a high-sensitivity
spectropolarimetric observation of the quiet solar chromosphere in
the strongest (8542 \AA) line of the \CaII~IR triplet. It was obtained
by R. Ramelli (IRSOL), R. Manso Sainz (IAC) and me using the
Z\"urich Imaging Polarimeter (ZIMPOL) attached to THEMIS. The
observed Stokes $V/I$ profiles are clearly caused by the longitudinal
Zeeman effect, while the Stokes $Q/I$ and $U/I$ signals are produced
mainly by the influence of atomic level
polarization. As seen in Fig.~\ref{jtb-fig:themis-zimpol}, although
the spatio-temporal resolution of this spectropolarimetric observation
is rather low (i.e., no better than $3\arcsec$ and 20 minutes), the
fractional polarization amplitudes fluctuate between 0.01\% and 0.1\%
along the spatial direction of the spectrograph's slit, with a typical
spatial scale of $5\arcsec$. Interestingly enough, while the Stokes
$Q/I$ signal changes its amplitude but remains always positive along
that spatial direction, the sign of the Stokes $U/I$ signal
fluctuates.

\begin{figure}
\centering
 \includegraphics[width=11cm]{\figspath/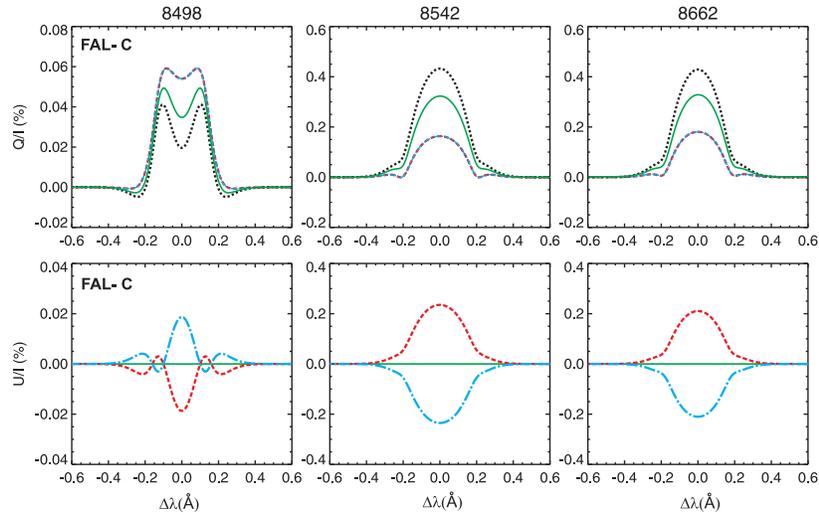}
 \caption{The emergent fractional linear polarization of the \CaII~IR
 triplet calculated for a LOS with $\mu=0.1$ in a semi-empirical model
 of the solar atmosphere. For the $B=0$ case (dotted curves)
 only $Q/I$ is non-zero. The other curves refer to a horizontal 5~mG
 magnetic field pointing towards the observer (red dashed curves),
 perpendicular to the observer (green solid curves) and away from
 the observer (blue dash-dotted curves). The reference direction for Stokes
 $Q$ is the parallel to the closest limb. Note that while the $Q/I$ amplitude
 fluctuates but remains always positive, the sign of $U/I$ depends on
 the azimuth of the magnetic field. From \citet{manso-trujillo-09}.  }
\label{jtb-fig:ca-IR-falc}
\end{figure}

The physical interpretation of this type of spectropolarimetric
observations requires solving the non-LTE problem of the $2^{\rm nd}$
kind for the \CaII~IR triplet. Fig.~\ref{jtb-fig:ca-IR-falc} shows
examples of the emergent fractional linear polarization calculated
with MULTIPOL in a semi-empirical model of the solar atmosphere. The
top panels show the emergent $Q/I$ profiles for a LOS with $\mu=0.1$,
for the unmagnetized reference case and for three possible
orientations of a 5~mG horizontal magnetic field. The bottom panels
show the corresponding $U/I$ signals, which are of course zero for the
unmagnetized case. Note that while the amplitudes of the theoretical
$Q/I$ profiles change with the strength and orientation of the
magnetic field and are always positive, the sign of $U/I$ is sensitive
to the azimuth of the magnetic field vector. Therefore, as expected,
the spatial variations in the observed fractional linear polarization
(see Fig.~\ref{jtb-fig:themis-zimpol}) are mainly due to changes in
the orientation of the chromospheric magnetic field.

These types of polarization signal resulting from atomic level
polarization and the Hanle and Zeeman effects can be exploited to
explore the thermal and magnetic structure of the solar
chromosphere. They can also be used to evaluate the degree of realism
of magneto-hydrodynamic simulations of the photosphere-chromosphere
system via careful comparisons of the observed Stokes profiles with
those obtained through forward-modeling calculations.

\section{Plasma structures in the chromosphere and corona}\label{jtb-sec5}

As mentioned above, a suitable diagnostic window for inferring the
magnetic field vector of plasma structures embedded in the solar
atmosphere is that provided by the polarization produced by the joint
action of atomic level polarization and the Hanle and Zeeman effects
in the \HeI~10830~\AA\ and 5876 \AA\ multiplets.

The resolved components of both helium multiplets stand out in
emission when observing off-limb structures at a given height above
the visible limb. Since their respective spectral lines have different
sensitivities to the Hanle effect, one would benefit from observing
them simultaneously in spicules and prominences. At present, such
simultaneous spectropolarimetric observations can be carried out with
THEMIS and with the polarimeter SPINOR attached to the Dunn Solar
Telescope. The main uncertainty with off-limb observations is that we
do not know whether the observed plasma structure was really in the
plane of the sky during the observing period (i.e., it is not known
whether the observed Stokes profiles were produced in 90$^{\circ}$
scattering geometry or not).

Concerning on-disk observations it is clear that the \HeI~10830~\AA\
triplet is the most suitable one, given that it shows significantly
more absorption than the \HeI~D$_3$ multiplet when observing a variety
of plasma structures against the bright background of the solar
disk. The additional fact that the Hanle effect in forward scattering
creates measurable linear polarization signals in the lines of the
\HeI~10830~\AA\ multiplet when the magnetic field is inclined with
respect to the local solar vertical direction
\citep{trujillo-nature02}, and that there is a nearby photospheric
line of \SiI , makes the 10830 \AA\ spectral region very suitable for
investigating the coupling between the photosphere and the corona. The
main uncertainty with on-disk observations is that we do not know the
exact height above the solar visible surface where the observed plasma
structure is located.

\subsection{Off-limb diagnostics of prominences and spicules}

The off-limb observational strategy consists in doing
spectropolarimetric observations with the image of the spectrograph's
slit at given, ideally consecutive distances from the visible solar
limb. The Stokes profiles measured at each pixel (or after downgrading
the original spatial resolution to increase the signal-to-noise ratio)
are then used to obtain information on the strength, inclination and
azimuth of the magnetic field vector via the application of Stokes
inversion techniques like those discussed in \S4.2. All published
applications to the \HeI~D$_3$ multiplet observations are based on the
optically-thin plasma assumption (e.g., \cite{landi-82},
\cite{casini-03}, \cite{ariste-casini-spicules}), while some of the
reported \HeI~10830~\AA\ observations were interpreted taking into
account radiative transfer effects through the constant-property slab
model discussed in \S4.2 (see \cite{trujillo-nature02},
\cite{trujillo-spicules} and \cite{hazel}).

For example, \citet{casini-03} interpreted new spectropolarimetric
observations of quiescent solar prominences in the \HeI~D$_3$
multiplet via the application of a PCA inversion code
(\cite{ariste-casini-pca}) based on an extensive database of
theoretical Stokes profiles. These profiles were calculated using the
optically thin approximation, covering all the magnetic configurations
and scattering geometries of interest. For one of the observed
prominences \citet{casini-03} provided two-dimensional maps of the
inferred magnetic field vector. Such maps showed the surprising
feature of areas reaching up to 80~G within a background of prominence
plasma with predominantly horizontal fields having
10--20~G (see also \cite{casini-bevilacqua}). 
\citet{paletou-01} and \citet{trujillo-nature02} also
inferred prominence fields significantly stronger than those found by
Leroy and collaborators (see \cite{leroy}). For a detailed
\HeI~10830~\AA\ investigation of the magnetic field vector in a
polar-crown prominence see \citet{merenda-06}.

Concerning spicules, it is important to emphasize that the observation
and theoretical modeling of the Hanle and Zeeman effects in such
spike-like jet features provides a suitable tool for investigating the
magnetism of the solar chromosphere. The paper by
\citet{centeno-spicules} discusses briefly the results that several
researchers have obtained through the interpretation of off-limb
observations and reports on a recent investigation based on the
application of the inversion code HAZEL to \HeI~10830~\AA\
spectropolarimetric observations of spicules in the quiet solar
chromosphere. They find magnetic fields with $B \approx 40$~G in a
localized area of the slit-jaw image, which could represent a possible
lower value for the field strength of organized network spicules.

\begin{figure}
\centering
 \includegraphics[angle=90,width=5.8cm]{\figspath/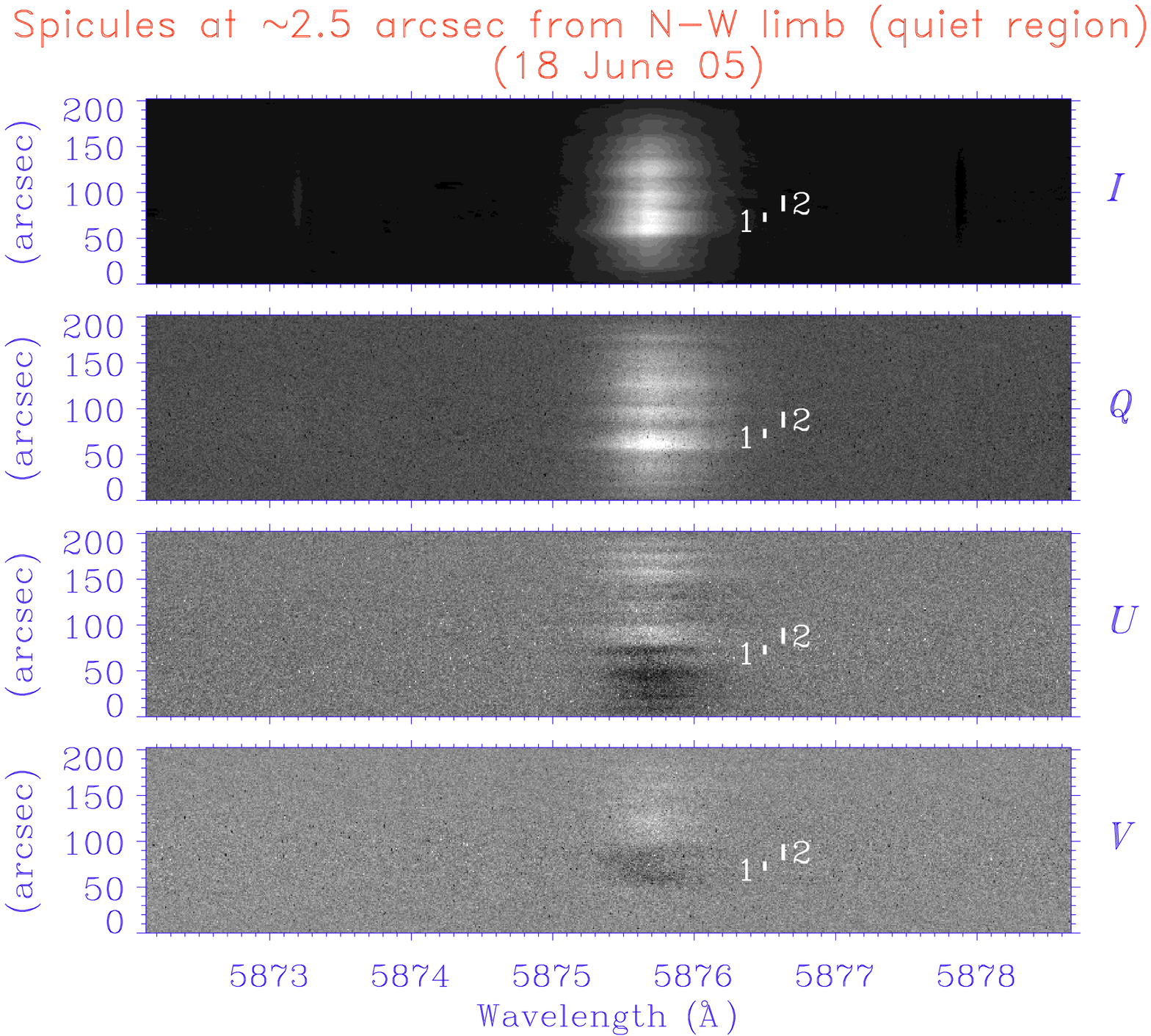}
 \includegraphics[angle=90,width=5.8cm]{\figspath/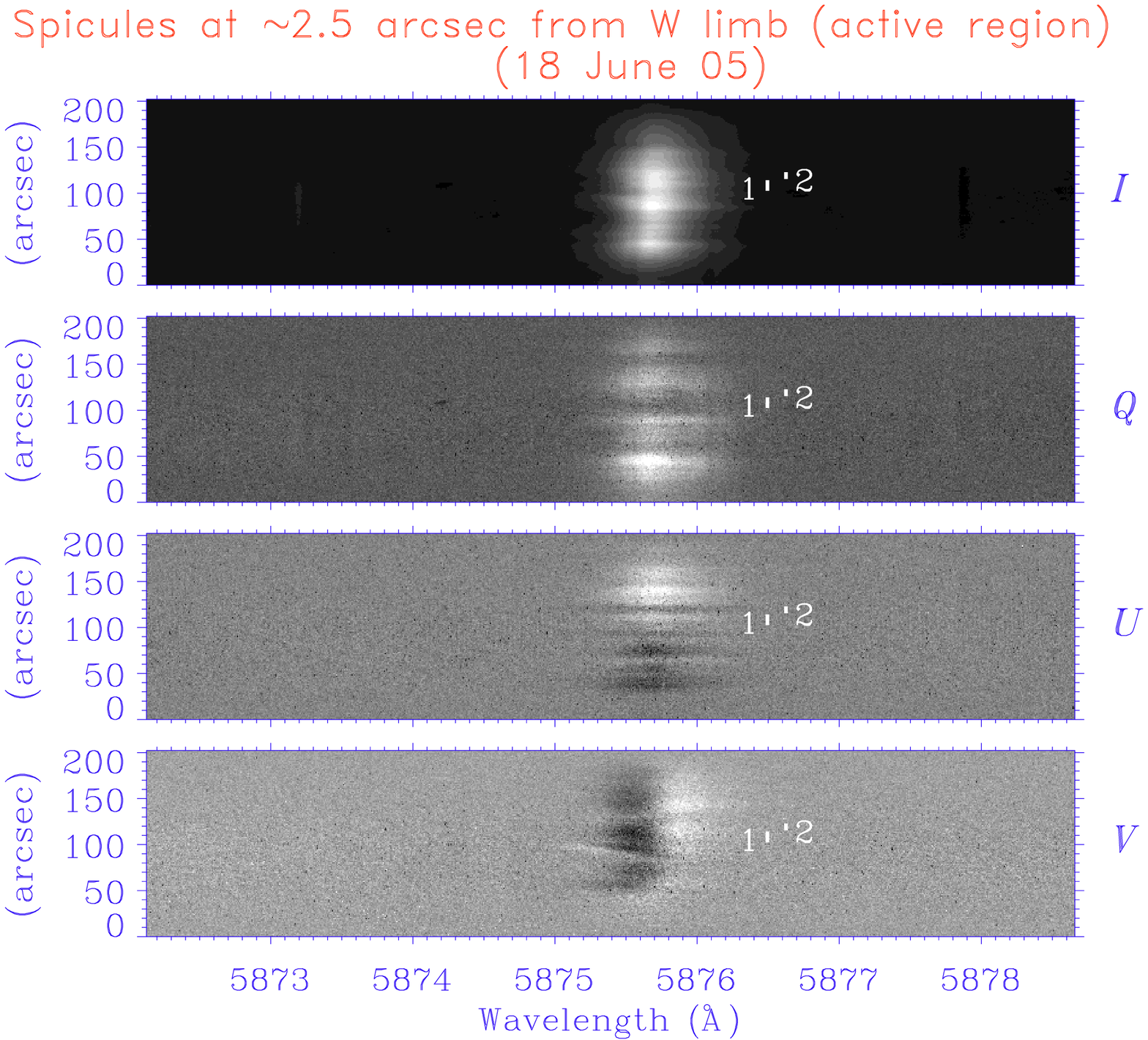}
 \caption{Illustrative examples of the off-limb Stokes profiles of the
 \HeI~5876~\AA\ multiplet observed in a quiet region (left panels) and
 close to an active region (right panels). Interestingly, the inferred
 magnetic field vector has $B \approx 10$~G in the quiet region, while
 a best fit to the profiles observed close to the active region
 requires $B \approx 50$~G. The reference direction for Stokes $Q$ is
 the parallel to the solar limb. From
\citet{ramelli-spicules}.}
  \label{jtb-fig:D3-spicules}
\end{figure}

Very interesting \HeI~D$_3$ observations of spicules are shown in
Fig.~\ref{jtb-fig:D3-spicules}.  The Stokes profiles of the left panel
correspond to a quiet region, while those of the right panel were
observed close to an active region. Obviously, for the magnetic
strengths of spicules (i.e., $B \lesssim 100$~G) the observed linear
polarization is fully dominated by the selective emission processes
that result from the atomic alignment of the upper levels of the
3d$^3$D term, without any significant influence of the transverse
Zeeman effect. Note that in both regions Stokes $U$ is non-zero, which
is the observational signature of the Hanle effect of an inclined
magnetic field. The change of sign in Stokes $U$ along the spatial
direction of the spectrograph's slit can be easily explained by
variations in the azimuth of the magnetic field vector. In contrast,
the circular polarization profiles of the D$_3$ multiplet are the
result of the joint action of the longitudinal Zeeman effect and of
atomic level orientation (see \S2). Interestingly, the
Stokes $V$ profiles corresponding to the observed quiet region are
dominated by atomic level orientation, while those 
observed in the spicules close to the active region are caused mainly
by the longitudinal Zeeman effect.

\subsection{On-disk diagnostics of filaments}

A significant difference between quiet region (QR) filaments and
active region (AR) filaments is that the former are weakly magnetized
(i.e., with $B<100$~G) and embedded in the $10^6$ K solar corona,
while the latter are strongly magnetized (i.e., with
$100\,{\rm G} \lesssim B \lesssim 1000$~G) and located at much lower heights
above the solar visible surface.
 
\begin{figure}  
  \centering
  \includegraphics[width=11.0cm]{\figspath/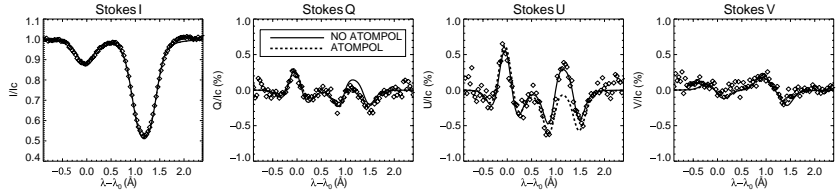}
\caption[]{
  Example of Stokes profiles dominated by the Zeeman effect observed
  in an AR filament.  Both best fits obtained with HAZEL give
  $B \approx 750$ G and $\theta_B=97^{\circ}$.  This figure 
  results from an ongoing collaboration with Y.~Katsukawa (NAOJ).
\label{jtb-fig:ar-filament}
}\end{figure}

For magnetic strengths $B \lesssim 100$~G the linear polarization of
the \HeI~10830~\AA\ triplet is fully dominated by the atomic level
polarization that is produced by anisotropic radiation pumping
(\cite{trujillo-nature02}). For stronger magnetic fields, the
contribution of the transverse Zeeman effect cannot be
neglected. However, in principle, the emergent linear polarization
should still show an important contribution caused by the presence of
atomic level polarization, even for the unfavorable case of low-lying
plasma structures having magnetic field strengths as large as 1000~G
(see Fig.~2 in Trujillo Bueno \& Asensio Ramos 2007; see also
Fig.~\ref{jtb-fig:atompol} of \S4.2).  Surprisingly, at the Fourth
International Workshop on Solar Polarization V.~Mart\'\i nez Pillet
and collaborators reported that the  Stokes $Q$ and $U$ profiles of the
\HeI~10830~\AA\ multiplet from an active-region filament
had the typical shape of polarization profiles produced by the
transverse Zeeman effect. Such observations of an AR filament on top
of a dense plage region at the polarity inversion line (i.e., an
``abutted'' plage region) have been recently analyzed in detail by
\citet{kuckein}, showing that the filament magnetic fields were mainly
horizontal and with strengths between 600 and
700~G. Figure~\ref{jtb-fig:ar-filament} shows another example of Stokes
profiles dominated by the Zeeman effect, observed in a (low-lying)
filament in an AR with large sunspots.

What is the explanation of this enigmatic finding?
According to \citet{trujillo-asensio-07}, the AR filament analyzed by
\citet{kuckein} had significant optical thickness so that the
radiation field generated by the structure itself reduced the positive
contribution to the anisotropy factor caused by the radiation from the
underlying solar photosphere (see Fig.~4 of
\cite{trujillo-asensio-07}). At meaningful optical thickness in the
``horizontal" direction (e.g., along the filament axis), the amount of
atomic level alignment in the filament can be significantly reduced,
to the extent that the transverse Zeeman effect of its {\em strong}
horizontal field dominates the emergent linear polarization (e.g., a
vertical light beam of intensity $I_c$ and two oppositely directed
horizontal beams of intensity $I_{\rm filament}=S=I_c/4$ produce {\em
zero anisotropy} in the horizontal reference system). An alternative
possibility, suggested by \citet{casini-ms-low}, requires the presence
of a randomly oriented field entangled with the main filament field,
and of similar magnitude (700~G!).

\begin{figure}  
  \centering
  \includegraphics[width=11.0cm]{\figspath/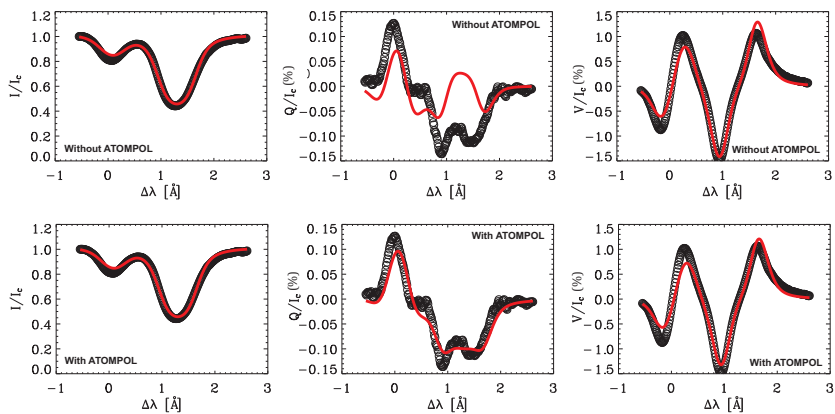}
\caption[]{
  Example of the Stokes profiles observed in an AR filament a few hours
  before its eruption, using the Tenerife Infrared Polarimeter 
  attached to the German VTT. 
  {\em Upper panels\/}: the best theoretical fit
  obtained with HAZEL when neglecting atomic level polarization
  corresponds to $B=460$~G. {\em Lower panels\/}: the best theoretical
  fit when atomic level polarization is taken into account corresponds
  to $B=390$~G. In both cases the inferred inclination is
  $\theta_B=115^{\circ}$. The Stokes $U$ profile (not shown
  here) is more or less similar to the observed Stokes $Q$, but of
  opposite sign.
\label{jtb-fig:filament-flare}
}\end{figure}

Another important question is whether all AR filaments permeated by a
significantly strong (e.g., $B \approx 500$~G) and predominantly 
horizontal magnetic field show linear polarization profiles dominated
by the transverse Zeeman effect. The fact that the answer to this
question is {\em negative\/} can be seen in
Fig.~\ref{jtb-fig:filament-flare}, which shows an example of the
Stokes profiles that A.~Asensio Ramos (IAC), C.~Beck (IAC) and I
observed on June 9, 2007 in an AR filament several hours before its
eruption. The solid curves in the upper panels show the best
theoretical fits to the observed Stokes profiles that the inversion
code HAZEL gives when only the Zeeman effect is considered, while the
lower panels demonstrate that the fit to the observed linear
polarization is dramatically improved when the influence of atomic
level polarization is also taken into account. It is
interesting to note that, while the inferred inclination of the
magnetic field with respect to the local vertical is
$\theta_B \approx 115^{\circ}$ in both cases, the magnetic field
strength turns out to be 70~G stronger when Stokes inversion is
performed neglecting atomic level polarization.

As seen in Fig.~\ref{jtb-fig:filament-flare} the intensity profiles
observed in this AR filament show significant absorption, which
implies that the slab's optical thickness along the LOS needed to fit
them with HAZEL is significant. Therefore, following
\citet{trujillo-asensio-07}, one may argue that the anisotropy of the
radiation field within the filament was small and that, accordingly,
the emergent linear polarization should be dominated by the transverse
Zeeman effect. However, this was not the case (see the Stokes $Q$
panels of Fig.~\ref{jtb-fig:filament-flare}). I believe that the
solution to this enigma has to do with the degree of compactness of
the multitude of individual magnetic fibrils, stacked one upon
another, that together determine the structure of a filament. Very
likely, the plasma of the AR filament analyzed by \citet{kuckein} had
a high degree of compactness, such that a single constant-property
slab or tube model with a significant optical thickness in the
horizontal direction provides a suitable representation. As a result,
the radiation field generated by the plasma structure itself may have
produced a negative contribution to the anisotropy factor, so that the
anisotropy of the true radiation field that illuminated the helium
atoms in the filament body was significantly smaller than that
corresponding to the optically thin case (see Fig.~4 of
\cite{trujillo-asensio-07}).  On the contrary, in my opinion, a more
suitable model for the AR filament we observed on June 9, 2007 a few
hours before its eruption is that of a multitude of {\em optically
thin} threads of magnetized plasma, each of them illuminated by the
anisotropic radiation coming from the underlying photosphere, and such
that the total optical thickness along the LOS is that needed by HAZEL
to fit the observed Stokes $I(\lambda)$ profiles. The observational
and theoretical support in favour of prominence thread structure is
overwhelming (e.g., the review by \cite{heinzel}), but it is
interesting to note that the signature of atomic level polarization in
the linear polarization profiles observed in some AR filaments may
provide information on the degree of compactness of the structure that
results from the agglomeration of multitude of individual magnetized
fibrils.

Finally, it is of interest to mention that on 11 September 2003
\citet{merenda-coimbra} carried out \HeI~spectropolarimetric
observations of a filament in a moderately active region that was
located relatively close to the solar disk center, and found that the
Stokes $Q$ and $U$ profiles were dominated by the influence of atomic
level polarization in most of the filament body, except in a small
region of the filament apparently close to one of its footpoints
(where the observed linear polarization clearly resulted from the
joint action of atomic level polarization and the transverse Zeeman
effect). A detailed interpretation of these observations via the
Stokes inversion strategy described in \citet{merenda-06} allowed us
to infer a full magnetic map of the filament, with the azimuth,
inclination and strength of the magnetic field vector at each point
within the filament (see \cite{merenda-coimbra} and
\cite{merenda-thesis}). Interestingly, while the inferred magnetic
field vector was predominantly horizontal, practically aligned with
the local filament axis and with $B \approx 100$~G in most points of
the filament body, it was found to be significantly stronger (a few
hundred gauss) in the part of the filament apparently close to one of
its footpoints.

\subsection{Magnetic field ``reconstruction" in emerging flux regions}

\begin{figure}  
  \centering
  \includegraphics[width=11cm]{\figspath/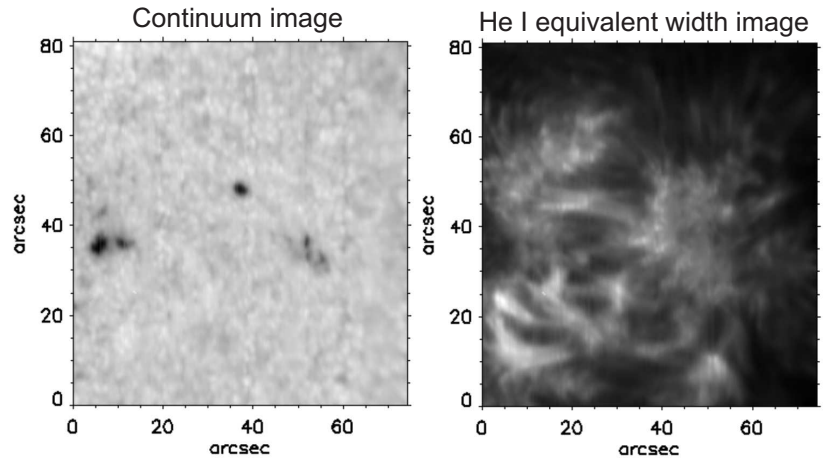}
\caption[]{
  Observation of a region with small pores obtained in collaboration
  with M. Collados (IAC) using the Tenerife Infrared Polarimeter
  attached to the VTT.
\label{jtb-fig:pores}
}\end{figure}

Figure~\ref{jtb-fig:pores} shows results 
from spectropolarimetric measurements of
a region with small pores observed in the 10830 \AA\
spectral region. The left panel is a continuum image of the observed
field of view, while the bright features in the right panel show the
pixels where the equivalent width of the Stokes $I(\lambda)$ profile
of the \HeI~10830~\AA\ red blended component was significant. This is
one of the most spectacular \HeI~10830~\AA\ equivalent width images I
have ever seen. It shows loop-like structures in an emerging flux
region with the largest equivalent width values localized at the apex
of the loops.

It is not obvious to identify loops in the equivalent width image of
the region observed by \citet{solanki-03}. In a recently published
research note \citet{judge} has pointed out that if there were really
loops in the region analyzed by \citet{solanki-03} they must have been
near the plane containing the loop footpoints and the LOS, otherwise
their loop images would have appeared significantly curved. The
conclusion by \citet{solanki-03} that their data demonstrate the
presence of a current sheet {\em in the upper solar atmosphere\/} is
based on the assumption that the observed absorption in the
\HeI~10830~\AA\ triplet forms along magnetic field lines, which
allowed them to associate a formation height to the magnetic field
vector inferred at each pixel via the application of their
Milne-Eddington inversion code. \citet{judge} has criticized this
assumption on which the magnetic field ``reconstruction'' technique of
\citet{solanki-03} is based. He argues that \HeI~10830~\AA\ formation
along a horizontal slab is a more reasonable assumption that along the
skin of the emerging flux region and that the magnetic field
reconstructions by \citet{solanki-03} lead to spurious results. I
think, however, that the \HeI~equivalent width image of
Fig.~\ref{jtb-fig:pores} supports Solanki et al.'s assumption that in
regions of emerging flux the \HeI~triplet is formed within loops, but
it is not clear to me whether ``freshly emerged loops'' were really
present in the region they observed. Fortunately, it seems clear that
loops were really present in the region corresponding to
Fig.~\ref{jtb-fig:pores}. Hence, we will soon try to invert the
ensuing data with HAZEL in order to check carefully if the application
of the reconstruction technique of \citet{solanki-03} leads to loops
coinciding with those seen in Fig.~\ref{jtb-fig:pores}. This is
important because, as far as I know, the magnetic field reconstruction
technique proposed by \citet{solanki-03} is the only available to
deliver three-dimensional information about the magnetic field
vector from \HeI~10830~\AA\ spectropolarimetry.

\section{How to map the magnetic fields of coronal loops ?}

The primary emission of the $10^6$ K solar coronal plasma lies in the
EUV and soft X-rays, two spectral regions that can only be observed
from space. At such short wavelengths the coronal magnetic fields are
unable to produce any significant Zeeman polarization
signal. Therefore, a key question is: can we expect scattering
polarization in permitted lines at EUV wavelengths for which the
underlying quiet solar disk is seen completely dark?
\citet{manso-trujillo-spw5} think that the answer is yes for the
following reasons. For some EUV lines their lower level is the ground
level of a forbidden line at visible or near-IR wavelengths, which is
polarized due to the anisotropic illumination of the atoms at the
forbidden line wavelength. This lower-level atomic alignment is
transferred to the upper level of the EUV line by collisional
excitation. Therefore, since the upper level of the EUV line can be
polarized, we may have measurable scattering polarization signals
caused by the ensuing selective emission processes in the allowed EUV
line. The linear polarization thus generated would be sensitive to the
electronic density ($N_{\rm e}$) and to the orientation of the
magnetic field vector, although not to its strength because for
permitted EUV lines $B \gg B_{\rm H}(J_l)$ and $B \ll B_{\rm H}(J_u)$
(with $B$ the magnetic strength in the coronal plasma). Interestingly
enough, contrary to the case of forbidden line polarimetry (e.g.,
\cite{casini-judge}, \cite{tomczyk}), such linear polarization in
allowed EUV transitions would be observable also in forward scattering
at the solar disk center. This is extremely important because it
provides a way for mapping the magnetic field of the extended solar
atmosphere all the way up from the photosphere to the corona.

There are several interesting EUV lines satisfying these requirements.
For example, the theoretical prediction of \citet{manso-trujillo-spw5}
for the Fe~X line at 174.5~\AA\ is that the ensuing $Q/I$ amplitude in
$90^{\circ}$ scattering geometry at a height of 0.1 solar radii above
the solar surface varies between about $0.1\%$ for $N_{\rm
e}=10^9$~cm$^{-3}$ and $5\%$ for $N_{\rm e}=10^7$~cm$^{-3}$, being a
factor two smaller for the case of a horizontal magnetic field
observed against the solar disk in forward-scattering geometry.

\section{Concluding comments}\label{jtb-conclusions}

We should put high-sensitivity spectropolarimeters on ground-based and
space telescopes for simultaneously measuring the polarization in
photospheric and chromospheric lines.  A very good choice would be the
spectral region of the IR triplet of \CaII~and/or that of the
\HeI~10830~\AA\ triplet. If, in addition, we want to do something
technologically challenging, we could then put a EUV imaging
polarimeter on a space telescope (i.e., a TRACE-like instrument, but
capable of obtaining also linear polarization images of coronal
loops).

It would also be of great scientific interest to put a
high-sensitivity spectropolarimeter in space in order to simply {\em
discover\/} what the linearly polarized UV spectrum of the Sun looks
like. For example, lines like \MgII\,k at 2795~\AA\ and
the Ly\,${\alpha}$ and Ly\,${\beta}$ lines of hydrogen are expected to
show measurable linear polarization signals, also when pointing at
the solar disk center where we observe the forward scattering case
(\cite{trujillo-cosmic-vision}). Such observations would provide
precious information on the magnetic field structuring of the solar
transition region from the chromosphere to the $10^6$ K solar coronal
plasma.

Concerning improvements in the diagnostic tools of chromospheric and
coronal fields, our next step will be to acknowledge that ``the Sun is
a wolf in sheep's clothing'' and to generalize the methods reported
here to consider more sophisticated radiative transfer models and to
account for variations of the magnetic fields and flows at
sub-resolution scales. In fact, although many of the Stokes profiles
observed in chromospheric lines can be fitted with one-component
models, some of the observations discussed in this review and many
others (e.g., \cite{sn-tb-rc-science}, \cite{centeno-05},
\cite{lagg-07}, \cite{sasso-coimbra}) indicate the need to consider at
least two magnetic field components with different flows and/or
orientations.

\begin{acknowledgement}
  It is a pleasure to thank my colleagues and friends from Tenerife,
  Locarno, Firenze, Boulder and Tokyo for fruitful discussions and
  collaborations on the topic of this review.  The author is also
  grateful to the editors and the other SOC members of the
  Evershed centenary meeting for their invitation to participate in a
  very good conference.  Financial support by the Spanish Ministry of
  Science through project AYA2007-63881 and by the European Commission
  via the SOLAIRE network (MTRN-CT-2006-035484) are gratefully
  acknowledged.
\end{acknowledgement}

\begin{small}


\bibliographystyle{rr-assp}       


\end{small}

\end{document}

%% file: rr-assp-defs.tex

\def\thisvolume{these proceedings}

\def\aj{{AJ}}			
\def\araa{{ARA\&A}}		
\def\apj{{ApJ}}			
\def\apjl{{ApJ}}		
\def\apjs{{ApJS}}		
\def\ao{{Appl.\ Optics}} 
\def\apss{{Ap\&SS}}		
\def\aap{{A\&A}}		
\def\aapr{{A\&A~Rev.}}		
\def\aaps{{A\&AS}}		
\def\an{{Astron.\ Nachrichten}}
\def\aspcs{{ASP Conf.\ Ser.}}
\def\assp{{Astrophys.\ \& Space Sci.\ Procs., Springer, Heidelberg}}
\def\azh{{AZh}}			
\def\baas{{BAAS}}		
\def\jrasc{{JRASC}}	
\def\memras{{MmRAS}}		
\def\mnras{{MNRAS}}
\def\nat{{Nat}}		
\def\pra{{Phys.\ Rev.\ A}} 
\def\prb{{Phys.\ Rev.\ B}}		
\def\prc{{Phys.\ Rev.\ C}}		
\def\prd{{Phys.\ Rev.\ D}}		
\def\prl{{Phys.\ Rev.\ Lett.}} 
\def\pasp{{PASP}}
\def\pasj{{PASJ}}		
\def\qjras{{QJRAS}}
\def\science{{Sci}}		
\def\skytel{{S\&T}}		
\def\solphys{{Solar\ Phys.}} 
\def\sovast{{Soviet\ Ast.}}  
\def\ssr{{Space\ Sci.\ Rev.}}
\def\svassp{{Astrophys.\ Space Sci.\ Procs., Springer, Heidelberg}}
\def\zap{{ZAp}}			
\let\astap=\aap
\let\apjlett=\apjl
\let\apjsupp=\apjs
\def\grl{{Geophys.\ Res.\ Lett.}}  
\def\jgr{{J. Geophys.\ Res.}} 

\def\ion#1#2{{\rm #1}\,{\uppercase{#2}}}  
\def\deg{\hbox{$^\circ$}}
\def\sun{\hbox{$\odot$}}
\def\earth{\hbox{$\oplus$}}
\def\la{\mathrel{\hbox{\rlap{\hbox{\lower4pt\hbox{$\sim$}}}\hbox{$<$}}}}
\def\ga{\mathrel{\hbox{\rlap{\hbox{\lower4pt\hbox{$\sim$}}}\hbox{$>$}}}}
\def\sq{\hbox{\rlap{$\sqcap$}$\sqcup$}}
\def\arcmin{\hbox{$^\prime$}}
\def\arcsec{\hbox{$^{\prime\prime}$}}
\def\fd{\hbox{$.\!\!^{\rm d}$}}
\def\fh{\hbox{$.\!\!^{\rm h}$}}
\def\fm{\hbox{$.\!\!^{\rm m}$}}
\def\fs{\hbox{$.\!\!^{\rm s}$}}
\def\fdg{\hbox{$.\!\!^\circ$}}
\def\farcm{\hbox{$.\mkern-4mu^\prime$}}
\def\farcs{\hbox{$.\!\!^{\prime\prime}$}}
\def\fp{\hbox{$.\!\!^{\scriptscriptstyle\rm p}$}}
\def\micron{\hbox{$\mu$m}}
\def\onehalf{\hbox{$\,^1\!/_2$}}	
\def\onethird{\hbox{$\,^1\!/_3$}}
\def\twothirds{\hbox{$\,^2\!/_3$}}
\def\onequarter{\hbox{$\,^1\!/_4$}}
\def\threequarters{\hbox{$\,^3\!/_4$}}
\def\ubv{\hbox{$U\!BV$}}		
\def\ubvr{\hbox{$U\!BV\!R$}}		
\def\ubvri{\hbox{$U\!BV\!RI$}}		
\def\ubvrij{\hbox{$U\!BV\!RI\!J$}}		
\def\ubvrijh{\hbox{$U\!BV\!RI\!J\!H$}}		
\def\ubvrijhk{\hbox{$U\!BV\!RI\!J\!H\!K$}}		
\def\ub{\hbox{$U\!-\!B$}}		
\def\bv{\hbox{$B\!-\!V$}}		
\def\vr{\hbox{$V\!-\!R$}}		
\def\ur{\hbox{$U\!-\!R$}}


\def\labelitemi{{\bf --}}  

\def\rmit#1{{\it #1}}              
\def\rmit#1{{\rm #1}}              
\def\etal{\rmit{et al.}}           
\def\etc{\rmit{etc.}}           
\def\ie{\rmit{i.e.,}}              
\def\eg{\rmit{e.g.,}}              
\def\cf{cf.}                       
\def\viz{\rmit{viz.}}
\def\vs{\rmit{vs.}}

\def\rot{\hbox{\rm rot}}
\def\div{\hbox{\rm div}}
\def\lesssim{\mathrel{\hbox{\rlap{\hbox{\lower4pt\hbox{$\sim$}}}\hbox{$<$}}}}
\def\gtrsim{\mathrel{\hbox{\rlap{\hbox{\lower4pt\hbox{$\sim$}}}\hbox{$>$}}}}
\def\mathstacksym#1#2#3#4#5{\def#1{\mathrel{\hbox to 0pt{\lower 
    #5\hbox{#3}\hss} \raise #4\hbox{#2}}}}
\mathstacksym\lesssim{$<$}{$\sim$}{1.5pt}{3.5pt} 
\mathstacksym\gtrsim{$>$}{$\sim$}{1.5pt}{3.5pt} 
\mathstacksym\lrarrow{$\leftarrow$}{$\rightarrow$}{2pt}{1pt} 
\mathstacksym\lessgreat{$>$}{$<$}{3pt}{3pt} 

\def\dif{\: {\rm d}}                       
\def\ep{\:{\rm e}^}                        
\def\dash{\hbox{$\,-\,$}}                  
\def\is{\!=\!}                             

\def\starname#1#2{${#1}$\,{\rm {#2}}}  
\def\Teff{\hbox{$T_{\rm eff}$}}   

\def\kms{\hbox{km$\;$s$^{-1}$}}
\def\ms{\hbox{m$\;$s$^{-1}$}}
\def\Mxcm{\hbox{Mx\,cm$^{-2}$}}    

\def\Bapp{\hbox{$B_{\rm app}$}}    

\def\komega{($k, \omega$)}                 
\def\kf{($k_h,f$)}                         
\def\VminI{\hbox{$V\!\!-\!\!I$}}           
\def\IminI{\hbox{$I\!\!-\!\!I$}}           
\def\VminV{\hbox{$V\!\!-\!\!V$}}           
\def\Xt{\hbox{$X\!\!-\!t$}}                

\def\level #1 #2#3#4{$#1 \: ^{#2} \mbox{#3} ^{#4}$}   

\def\specchar#1{\uppercase{#1}}    
\def\AlI{\mbox{Al\,\specchar{i}}}  
\def\BI{\mbox{B\,\specchar{i}}} 
\def\BII{\mbox{B\,\specchar{ii}}}  
\def\BaI{\mbox{Ba\,\specchar{i}}}  
\def\BaII{\mbox{Ba\,\specchar{ii}}} 
\def\CI{\mbox{C\,\specchar{i}}} 
\def\CII{\mbox{C\,\specchar{ii}}} 
\def\CIII{\mbox{C\,\specchar{iii}}} 
\def\CIV{\mbox{C\,\specchar{iv}}} 
\def\CaI{\mbox{Ca\,\specchar{i}}} 
\def\CaII{\mbox{Ca\,\specchar{ii}}} 
\def\CaIII{\mbox{Ca\,\specchar{iii}}} 
\def\CoI{\mbox{Co\,\specchar{i}}} 
\def\CrI{\mbox{Cr\,\specchar{i}}} 
\def\CriI{\mbox{Cr\,\specchar{ii}}} 
\def\CsI{\mbox{Cs\,\specchar{i}}} 
\def\CsII{\mbox{Cs\,\specchar{ii}}} 
\def\CuI{\mbox{Cu\,\specchar{i}}} 
\def\FeI{\mbox{Fe\,\specchar{i}}} 
\def\FeII{\mbox{Fe\,\specchar{ii}}} 
\def\FeIX{\mbox{Fe\,\specchar{ix}}}
\def\FeX{\mbox{Fe\,\specchar{x}}}
\def\FeXVI{\mbox{Fe\,\specchar{xvi}}}
\def\FrI{\mbox{Fr\,\specchar{i}}}
\def\HI{\mbox{H\,\specchar{i}}} 
\def\HII{\mbox{H\,\specchar{ii}}} 
\def\Hmin{\hbox{\rmH$^{^{_{\scriptstyle -}}}$}}      
\def\Hemin{\hbox{{\rm He}$^{^{_{\scriptstyle -}}}$}} 
\def\HeI{\mbox{He\,\specchar{i}}} 
\def\HeII{\mbox{He\,\specchar{ii}}} 
\def\HeIII{\mbox{He\,\specchar{iii}}} 
\def\KI{\mbox{K\,\specchar{i}}} 
\def\KII{\mbox{K\,\specchar{ii}}} 
\def\KIII{\mbox{K\,\specchar{iii}}} 
\def\LiI{\mbox{Li\,\specchar{i}}} 
\def\LiII{\mbox{Li\,\specchar{ii}}} 
\def\LiIII{\mbox{Li\,\specchar{iii}}} 
\def\MgI{\mbox{Mg\,\specchar{i}}} 
\def\MgII{\mbox{Mg\,\specchar{ii}}} 
\def\MgIII{\mbox{Mg\,\specchar{iii}}} 
\def\MnI{\mbox{Mn\,\specchar{i}}} 
\def\NI{\mbox{N\,\specchar{i}}}
\def\NIV{\mbox{N\,\specchar{iv}}}
\def\NaI{\mbox{Na\,\specchar{i}}}
\def\NaII{\mbox{Na\,\specchar{ii}}}
\def\NaIII{\mbox{Na\,\specchar{iii}}}
\def\NeVIII{\mbox{Ne\,\specchar{viii}}} 
\def\NiI{\mbox{Ni\,\specchar{i}}} 
\def\NiII{\mbox{Ni\,\specchar{ii}}}
\def\NiIII{\mbox{Ni\,\specchar{iii}}} 
\def\OI{\mbox{O\,\specchar{i}}} 
\def\OVI{\mbox{O\,\specchar{vi}}}
\def\RbI{\mbox{Rb\,\specchar{i}}} 
\def\SII{\mbox{S\,\specchar{ii}}} 
\def\SiI{\mbox{Si\,\specchar{i}}} 
\def\SiII{\mbox{Si\,\specchar{ii}}} 
\def\SrI{\mbox{Sr\,\specchar{i}}}
\def\SrII{\mbox{Sr\,\specchar{ii}}}
\def\TiI{\mbox{Ti\,\specchar{i}}} 
\def\TiII{\mbox{Ti\,\specchar{ii}}} 
\def\TiIII{\mbox{Ti\,\specchar{iii}}} 
\def\TiIV{\mbox{Ti\,\specchar{iv}}} 
\def\VI{\mbox{V\,\specchar{i}}} 
\def\HtwoO{\mbox{H$_2$O}}        
\def\Otwo{\mbox{O$_2$}}          

\def\Halpha{\mbox{H\hspace{0.1ex}$\alpha$}} 
\def\Ha{\mbox{H\hspace{0.2ex}$\alpha$}}
\def\Hbeta{\mbox{H\hspace{0.2ex}$\beta$}}
\def\Hgamma{\mbox{H\hspace{0.2ex}$\gamma$}}
\def\Hdelta{\mbox{H\hspace{0.2ex}$\delta$}}
\def\Hepsilon{\mbox{H\hspace{0.2ex}$\epsilon$}}
\def\Hzeta{\mbox{H\hspace{0.2ex}$\zeta$}}
\def\Lyalpha{\mbox{Ly$\hspace{0.2ex}\alpha$}}
\def\Lybeta{\mbox{Ly$\hspace{0.2ex}\beta$}}
\def\Lygamma{\mbox{Ly$\hspace{0.2ex}\gamma$}}
\def\Lycont{\mbox{Ly\hspace{0.2ex}{\small cont}}}
\def\Baalpha{\mbox{Ba$\hspace{0.2ex}\alpha$}}
\def\Babeta{\mbox{Ba$\hspace{0.2ex}\beta$}}
\def\Bacont{\mbox{Ba\hspace{0.2ex}{\small cont}}}
\def\Paalpha{\mbox{Pa$\hspace{0.2ex}\alpha$}}
\def\Bralpha{\mbox{Br$\hspace{0.2ex}\alpha$}}

\def\NaD{\mbox{Na\,\specchar{i}\,D}}    
\def\NaDone{\mbox{Na\,\specchar{i}\,\,D$_1$}}
\def\NaDtwo{\mbox{Na\,\specchar{i}\,\,D$_2$}}
\def\NaID{\mbox{Na\,\specchar{i}\,\,D}}
\def\NaIDone{\mbox{Na\,\specchar{i}\,\,D$_1$}}
\def\NaIDtwo{\mbox{Na\,\specchar{i}\,\,D$_2$}}
\def\Done{\mbox{D$_1$}}
\def\Dtwo{\mbox{D$_2$}}

\def\Mgbone{\mbox{Mg\,\specchar{i}\,b$_1$}}
\def\Mgbtwo{\mbox{Mg\,\specchar{i}\,b$_2$}}
\def\Mgbthree{\mbox{Mg\,\specchar{i}\,b$_3$}}
\def\MgIb{\mbox{Mg\,\specchar{i}\,b}}
\def\MgIbone{\mbox{Mg\,\specchar{i}\,b$_1$}}
\def\MgIbtwo{\mbox{Mg\,\specchar{i}\,b$_2$}}
\def\MgIbthree{\mbox{Mg\,\specchar{i}\,b$_3$}}

\def\CaIIK{\mbox{Ca\,\specchar{ii}\,K}}       
\def\CaIIH{\mbox{Ca\,\specchar{ii}\,H}}
\def\CaIIHK{\mbox{Ca\,\specchar{ii}\,H\,\&\,K}}
\def\HK{\mbox{H\,\&\,K}}
\def\Kthree{\mbox{K$_3$}}      
\def\Hthree{\mbox{H$_3$}}
\def\Ktwo{\mbox{K$_2$}}
\def\Htwo{\mbox{H$_2$}}
\def\Kone{\mbox{K$_1$}}     
\def\Hone{\mbox{H$_1$}}     
\def\KtwoV{\mbox{K$_{2V}$}}
\def\KtwoR{\mbox{K$_{2R}$}}
\def\KoneV{\mbox{K$_{1V}$}}
\def\KoneR{\mbox{K$_{1R}$}}
\def\HtwoV{\mbox{H$_{2V}$}}
\def\HtwoR{\mbox{H$_{2R}$}}
\def\HoneV{\mbox{H$_{1V}$}}
\def\HoneR{\mbox{H$_{1R}$}}

\def\hk{\mbox{h\,\&\,k}}
\def\kthree{\mbox{k$_3$}}    
\def\hthree{\mbox{h$_3$}}
\def\ktwo{\mbox{k$_2$}}
\def\htwo{\mbox{h$_2$}}
\def\kone{\mbox{k$_1$}}     
\def\hone{\mbox{h$_1$}}     
\def\ktwoV{\mbox{k$_{2V}$}}
\def\ktwoR{\mbox{k$_{2R}$}}
\def\koneV{\mbox{k$_{1V}$}}
\def\koneR{\mbox{k$_{1R}$}}
\def\htwoV{\mbox{h$_{2V}$}}
\def\htwoR{\mbox{h$_{2R}$}}
\def\honeV{\mbox{h$_{1V}$}}
\def\honeR{\mbox{h$_{1R}$}}

\ifnum\preprintheader=1     
\makeatletter  
\def\@maketitle{\newpage
\markboth{}{}%
  {\mbox{} \vspace*{-8ex} \par 
   \em \footnotesize To appear in ``Magnetic Coupling between the Interior 
       and the Atmosphere of the Sun'', eds. S.~S.~Hasan and R.~J.~Rutten, 
       Astrophysics and Space Science Proceedings, Springer-Verlag, 
       Heidelberg, Berlin, 2009.} \vspace*{-5ex} \par
 \def\lastand{\ifnum\value{@inst}=2\relax
                 \unskip{} \andname\
              \else
                 \unskip \lastandname\
              \fi}%
 \def\and{\stepcounter{@auth}\relax
          \ifnum\value{@auth}=\value{@inst}%
             \lastand
          \else
             \unskip,
          \fi}%
  \raggedright
 {\Large \bfseries\boldmath
  \pretolerance=10000
  \let\\=\newline
  \raggedright
  \hyphenpenalty \@M
  \interlinepenalty \@M
  \if@numart
     \chap@hangfrom{}
  \else
     \chap@hangfrom{\thechapter\thechapterend\hskip\betweenumberspace}
  \fi
  \ignorespaces
  \@title \par}\vskip .8cm
\if!\@subtitle!\else {\large \bfseries\boldmath
  \vskip -.65cm
  \pretolerance=10000
  \@subtitle \par}\vskip .8cm\fi
 \setbox0=\vbox{\setcounter{@auth}{1}\def\and{\stepcounter{@auth}}%
 \def\thanks##1{}\@author}%
 \global\value{@inst}=\value{@auth}%
 \global\value{auco}=\value{@auth}%
 \setcounter{@auth}{1}%
{\lineskip .5em
\noindent\ignorespaces
\@author\vskip.35cm}
 {\small\institutename\par}
 \ifdim\pagetotal>157\p@
     \vskip 11\p@
 \else
     \@tempdima=168\p@\advance\@tempdima by-\pagetotal
     \vskip\@tempdima
 \fi
}
\makeatother     
\fi